# Coexistence of quantized and non-quantized geometric phases in quasi-one-dimensional systems without inversion symmetry

Yi-Xin Xiao, Zhao-Qing Zhang, C. T. Chan[†]

Department of Physics and Institute for Advanced Study,

The Hong Kong University of Science and Technology,

Clear Water Bay, Kowloon, Hong Kong, China

It is well known that inversion symmetry in one-dimensional (1D) systems leads to the quantization of the geometric Zak phase to values of either $0$ or $\pi$. When the system has particle-hole symmetry, this topological property ensures the existence of zero-energy interface states at the interface of two bulk systems carrying different Zak phases. In the absence of inversion symmetry, the Zak phase can take any value and the existence of interface states is not ensured. We show here that the situation is different when the unit cell contains multiple degrees of freedom and a hidden inversion symmetry exists in a subspace of the system. As an example, we consider a system of two Su-Schrieffer-Heeger (SSH) chains coupled by a coupler chain. Although the introduction of coupler chain breaks the inversion symmetry of the system, a certain hidden inversion symmetry ensures the existence of a decoupled $2\times 2$ SSH Hamiltonian in the subspace of the entire system and the two bands associated with this subspace have quantized Zak phases. These "quantized" bands in turn can provide topological boundary or interface states in such systems. Since the entire system has no inversion symmetry, the bulk-boundary correspondence may not hold exactly. The above is also true when next-nearest-neighbor hoppings are included. Our systems can be realized straightforwardly in systems such as coupled single-mode optical waveguides or coupled acoustic cavities.

# I. Introduction

Topological phases and topological matter have attracted a lot of attention in many fields of physics[1,2]. Many exotic topological phases are discovered or realized in recent years, and good examples include integer quantum Hall systems[3], quantum anomalous Hall systems[4], topological insulators[1,2], topological superconductors[2,5], and topological semimetals[6]. Time reversal symmetry and particle-hole symmetry have been used to classify the topological phases into different classes[5]. The study of topological matters has also been extended to the realm of classical waves including photonics[7–19], acoustics[20–23] and elastic waves[24–27].

One of the key features of topological matters is its robust interface states at the boundary between two phases characterized by different values of the topological invariant[28]. Although the topological invariant is defined for the bulk, its physical significance manifests itself in the boundary: the value of the invariant determines the number of interface states and this is called bulk-boundary correspondence[28–30]. We note that the topological invariant is robust to deformations that do not close the bulk gap. When a gap is closed and reopened upon the variation of some physical parameters, the bands associated with the gap are inverted and characterized by different values of the topological invariant. Spatially confined mid-gap states can then be created at the interface of two topologically distinct bulk systems. The simplest example of bulk-boundary correspondence can be found in various 1D systems[31–34]. The inversion symmetric Su-Schrieffer-Heeger (SSH) model is one of the most studied 1D models which possess nontrivial topological properties[33]. Its topological properties are characterized by a $Z_2$ invariant given by the geometric Zak phase, which is the Berry phase defined in 1D systems[35]. According to the bulk-boundary correspondence, the Zak phase (in unit of $\pi$) or winding number is equal to the number of zero-energy boundary modes[36].

In 1D periodic systems, the Zak phase can be defined due to the ring topology of the Brillouin zone[35]. In the presence of inversion symmetry, symmetry analysis of the Wannier functions reveals that the Zak phase can only be 0 or $\pi$[35]. The inversion symmetry in a 1D system implies the existence of an inversion center about which the system is mirror symmetric. If the system lacks inversion symmetry, the Zak phases are generally not quantized and can take any values[37]. One interesting question is whether quantized Zak phase can exist in the absence of inversion symmetry. In this work, we show that this is possible if the unit cell of a system contains multiple degrees of freedom and some hidden inversion symmetry exists in a subsystem which is decoupled from the rest of the Hamiltonian. In such a case, the bands in the subsystem can have quantized Zak phases although the whole system does not have inversion symmetry.

We take a simple model of two identical SSH chains coupled by a coupler chain as an example[33,38]. In the absence of inter-chain couplings, two SSH chains are decoupled and all bands have quantized Zak phases. The introduction of inter-chain coupling breaks the inversion symmetry of the whole system so that the bands should not have quantized Zak phases according to current understanding. However, if the two SSH chains are identical, we find that there exists a decoupled $2 \times 2$ SSH Hamiltonian in a subspace of the whole system and the associated two bands have quantized Zak phases due to a hidden inversion symmetry in the subsystem. This is true for both nearest-neighbor (NN) hoppings and next-nearest-neighbor (NNN) hoppings. Then a question naturally arises: Will the well-known bulk-boundary correspondence still hold in this "semi-topological" system? In order to answer this question, we have studied the topological boundary and interface states. In the case of NN hoppings, it is found that the bulk-boundary correspondence may not hold exactly due to the existence of other non-topological zero-energy boundary modes outside the $2 \times 2$ subspace. However, when NNN hoppings are included, the bulk-boundary correspondence holds exactly. The model we propose here is likely to be the simplest inversion-symmetry-broken system that can support quantized Zak phases and can be realized by using coupled sound cavities or coupled single-mode waveguide systems[14]. We have also considered the case of three coupled SSH chains and similar results are found. The generalization to the case of any number of coupled SSH chains such as the Lieb-like ribbon[40,41] is also discussed.

The paper is organized as follows. In section II, we first introduce a system of two coupled SSH chains. We then explicitly show that although inversion symmetry is absent in the system, when the two SSH chains are identical, a hidden inversion symmetry exists in a subsystem which can be described by a $2 \times 2$ SSH Hamiltonian. In section III, we show both numerically and analytically that the Zak phase is quantized for the two bands of the subsystem when there are no NNN hoppings. As for other bands, the Zak phase is not quantized duo the absence of inversion symmetry of the whole system. In Section IV, topological boundary/interface states and the bulk-boundary correspondence in this "semi-topological" system are investigated and discussed. In Section V, the case of next-nearest-neighbor hoppings are investigated. A generalization to multiple coupled SSH chains are discussed and a summary is given in section VI.

## II. Model Hamiltonian

It is well known that inversion symmetry in 1D system leads to the quantization of Zak phases [35]. In the absence of inversion symmetry, the Zak phases can assume any value [35]. We show here that the above is not true in general, particularly when the unit cell of a system has multiple degrees of freedom and contains certain symmetries.

As a simple example, we consider a system of two SSH chains coupled indirectly through another chain of "atoms" as shown in Fig. 1(a). There are five "atoms" in a unit cell, which are labeled by numbers 1 to 5. The hopping parameters are represented by connections between two sites and labeled as $t_1, s_1, t_2, s_2, r_1, r_2$. The on-site energies are assumed to be $\epsilon_5$ for the $5^{th}$ site and zero (without loss of generality) for all the other sites. The presence of inter-chain couplings $r_1$ and $r_2$ breaks the inversion symmetry of the system and none of the five bands in the system has quantized Zak phase in the general situation. However, when the two SSH chains are identical, meaning that $t_1 = t_2 = t$ and $s_1 = s_2 = s$, by using the rank-nullity theorem [42], we show below that there always exists a decoupled $2 \times 2$ SSH Hamiltonian in a subspace of the whole system, in which the Zak phases of the two bands are quantized even though the entire system lacks inversion symmetry.

The total Hamiltonian of the system consists of two parts: two identical SSH chains $H_0$ and inter-chain coupling part $H_1$,

$$H = H_0 + H_1$$

$$H_0 = \sum_{i,n=1,3} t c_{i,n}^\dagger c_{i,n+1} + s c_{i+1,n}^\dagger c_{i,n+1} + H.c.$$

$$H_1 = \sum_i \epsilon_5 c_{i,5}^\dagger c_{i,5} + \sum_i \left( r_1 c_{i,1}^\dagger c_{i,5} + r_2 c_{i,3}^\dagger c_{i,5} + H.c. \right)$$

(1)

The operator $c_{i,n}$ $(c_{i,n}^\dagger)$ annihilates (creates) a particle at the site $(i,n)$, where $i$ and $n$ label the unit cells and sites within each unit cell, respectively.

By using $c_{km} = \frac{1}{\sqrt{N}} \sum_j c_{j,m} e^{-ikj}$ and defining the five-component annihilation operator as $\psi_k = [c_{k1}, c_{k2}, c_{k3}, c_{k4}, c_{k5}]^T$, the Hamiltonian (1) can be rewritten in momentum space as $H = \sum_k \psi_k^\dagger \mathcal{H}(k) \psi_k$, where

$$\mathcal{H}(k) = \begin{pmatrix} 0 & t+se^{-ik} & 0 & 0 & r_1 \\ t+se^{ik} & 0 & 0 & 0 & 0 \\ 0 & 0 & 0 & t+se^{-ik} & r_2 \\ 0 & 0 & t+se^{ik} & 0 & 0 \\ r_1 & 0 & r_2 & 0 & \epsilon_5 \end{pmatrix}. \quad (2)$$

Here we have taken the lattice constant $a$ as the unit of length and set $a = 1$. For this quasi-1D system, we can rearrange all the sites in Fig. 1(a) onto a truly 1D chain as shown in Fig. 1(b) with the connection lines representing the hoppings between two sites, where the line colors and labels are consistent with those in Fig. 1(a) when $t_1 = t_2 = t$ and $s_1 = s_2 = s$. Each rectangular box in Fig. 1(b) encloses a unit cell. The absence of inversion symmetry can be observed straightforwardly in Fig. 1(b) or by symmetry analysis of the Hamiltonian.

Without the inter-chain coupling part, the system has two doubly degenerate SSH bands, i.e., $E_\pm(k) = \pm\sqrt{t^2 + s^2 + 2ts \cos k}$. By using the eigenstates of these degenerate bands, we can construct a similarity transformation and re-express the Hamiltonian in Eq. (2) as

$$h(k) = X^{-1}\mathcal{H}(k)X = \begin{pmatrix} E_+ & 0 & 0 & 0 & r_1\eta^* \\ 0 & E_+ & 0 & 0 & r_2\eta^* \\ 0 & 0 & E_- & 0 & -r_1\eta^* \\ 0 & 0 & 0 & E_- & -r_2\eta^* \\ r_1\eta & r_2\eta & -r_1\eta & -r_2\eta & \epsilon_5 \end{pmatrix}, \quad (3)$$

where

$$\eta = \frac{\rho}{\sqrt{2}}, \rho = \frac{t+se^{-ik}}{E_+} = e^{-i\phi(k)}, \quad (4)$$

with $\phi(k)$ being the polar angle of the complex function $t + se^{ik} = E_+ e^{i\phi(k)}$, and

$$X = \frac{1}{\sqrt{2}}\begin{pmatrix} \rho & 0 & -\rho & 0 & 0 \\ 1 & 0 & 1 & 0 & 0 \\ 0 & \rho & 0 & -\rho & 0 \\ 0 & 1 & 0 & 1 & 0 \\ 0 & 0 & 0 & 0 & \sqrt{2} \end{pmatrix}. \quad (5)$$

In this representation, there are two $2 \times 2$ identity blocks $E_+ I_2$ and $E_- I_2$ in $h(k)$, which are coupled through two interaction blocks $h_{int+} = (r_1\eta, r_2\eta)$ and $h_{int-} = (-r_1\eta, -r_2\eta)$. We call the rank of the interaction block $h_{int\pm}$ as the number of effective hopping channels $p$. Here we have $p = Rank(h_{int\pm}) = 1$. According to the rank-nullity theorem, the dimension of the null space of this interaction block is $2 - p$, where 2 means the number of columns of $h_{int\pm}$. Therefore, with inter-chain couplings, the degeneracy of eigen-energy $E_\pm$ reduces to $2 - p = 1$ [43]. Thus, although the interaction lifts the degeneracy of two SSH bands, one set of SSH bands remains intact. The above arguments are especially useful in bipartite systems[44]. From a more physical point of view, we see that since the degeneracy ($D = 2$) in each block ($E_\pm I_2$) is larger than the number ($p = 1$) of effective coupling channels to other parts of the system, we are guaranteed to have $D - p = 1$ fold eigenvalue of $E_\pm$ unaffected, i.e., a pair of SSH bands remains. Since the system is guaranteed to have a set of SSH bands, we can use another similarity transformation to block-diagonalize the Hamiltonian in Eq. (2) into the following form:

$$\mathcal{H}_{BD}(k) = U^{-1}\mathcal{H}(k)U = \begin{pmatrix} 0 & t+se^{-ik} & 0 & 0 & 0 \\ t+se^{ik} & 0 & 0 & 0 & 0 \\ 0 & 0 & 0 & t+se^{-ik} & \sqrt{r_1^2+r_2^2} \\ 0 & 0 & t+se^{ik} & 0 & 0 \\ 0 & 0 & \sqrt{r_1^2+r_2^2} & 0 & \epsilon_5 \end{pmatrix},$$

(6)

where

$$U = \frac{1}{\sqrt{r_1^2+r_2^2}} \begin{pmatrix} -r_2 & 0 & r_1 & 0 & 0 \\ 0 & -r_2 & 0 & r_1 & 0 \\ r_1 & 0 & r_2 & 0 & 0 \\ 0 & r_1 & 0 & r_2 & 0 \\ 0 & 0 & 0 & 0 & \sqrt{r_1^2+r_2^2} \end{pmatrix}.$$

(7)

In the block diagonalized Hamiltonian $\mathcal{H}_{BD}(k)$, the $2 \times 2$ block has exactly the same form as the SSH Hamiltonian as aforementioned. An interesting question is whether the Zak phases of the two SSH bands derived from the $2 \times 2$ block of Eq. (6) are quantized. If the answer is positive, the system then can be divided into two subsystems: one can be characterized by a topological invariant and the other is non-topological. Such a system can be called "semi-topological" and, as far as we know, it has not been studied in the literature and new understandings for Zak phases and topological properties are anticipated.

### III. Quantization of Zak phase in the $2 \times 2$ subsystem

We show below both analytically and numerically that the two SSH bands derived from the $2 \times 2$ block of Eq. (6) indeed have quantized Zak phases. Numerically, the Zak phase is calculated by using the formula[45]

$$\theta = -\text{Im} \log \prod_{m=0}^{M-1} \langle u(k_m)|u(k_{m+1})\rangle, \tag{8}$$

where we have discretized the loop from $k = -\pi$ to $k = \pi$ into $M$ segments, and $|u(k_m)\rangle$ is the five-component cell-periodic eigenvector of the Hamiltonian in Eq. (2) at Bloch momentum $k_m = -\pi + m\Delta k$ with $\Delta k = 2\pi/M$. The periodic gauge is applied, i.e., $|u(-\pi)\rangle = |u(\pi)\rangle$. The band structures and the Zak phases of the two SSH bands for three typical situations are shown in Fig. 2(a)-(c): (a) the intra-cell hopping $t$ is larger than the inter-cell hopping $s$; (b) $t = s$; (c) $t < s$, where the on-site energy for the $5^{\text{th}}$ site is taken to be $\epsilon_5 = 0$. It is interesting to find that the two middle bands colored in blue are indeed quantized and have different Zak phases when the bands are gapped as shown in Fig. 2(a) and (c). The closing of the gap at $t = s$ implies a topological phase transition in this subsystem. The Zak phases are not quantized for the other three bands shown in red, which are given by the $3 \times 3$ block, due to the absence of inversion symmetry. Thus, we have a coexistence of quantized and non-quantized Zak phases in a single system. Since all the sites are assumed to have zero on-site energies, the system is bipartite and has chiral symmetry[43], which leads to a flat band at $E = 0$. The properties of the flat band in such systems with PT symmetry have been studied recently[38]. The wavefunction of the flat band vanishes on the minority sublattice sites, i.e., sites 1 and 3 in each unit cell.

Below we give an analytical proof of the quantization of Zak phases in the subsystem. First, we want to point out some special properties of the Bloch wavefunctions for the five bands in our system. If we denote the old "atomic orbitals" used in Eq. (2) as $|i\rangle$ and the new basis used in Eq. (6) as $|i'\rangle$, where $i = 1,2,3,4,5$, they are related by the unitary operator $U$ given in Eq. (7), i.e., $|i'\rangle = \sum_{j=1}^{5} U_{ji}|j\rangle$. Thus, the $2 \times 2$ SSH block in Eq. (6) is constructed in the space spanned by the two hybridized "atomic orbitals" below:

$$|1'\rangle = \frac{1}{\sqrt{r_1^2+r_2^2}}(-r_2|1\rangle + r_1|3\rangle), |2'\rangle = \frac{1}{\sqrt{r_1^2+r_2^2}}(-r_2|2\rangle + r_1|4\rangle), \quad (9)$$

where $\langle i'|j'\rangle = \delta_{ij}, i,j = 1,2$, and $\delta_{ij}$ is the Kronecker delta function. Equation (9) tells us that the Bloch wavefunctions for the two "quantized" bands have opposite signs for two pairs of sites $(1,3)$ and $(2,4)$ in each unit cell. Equation (9) also implies that in this subspace all wavefunctions vanish at site 5. For the $3 \times 3$ block in Eq. (6) the basis are $|3'\rangle = \frac{1}{\sqrt{r_1^2+r_2^2}}(r_1|1\rangle + r_2|3\rangle), |4'\rangle = \frac{1}{\sqrt{r_1^2+r_2^2}}(r_1|2\rangle + r_2|4\rangle)$ and $|5'\rangle = |5\rangle$. Hence the Bloch wavefunctions of the corresponding three bands have the same sign for the pairs $(1,3)$ and $(2,4)$. Since there exists an effective hopping between $|3'\rangle$ and $|5'\rangle$, the wavefunctions in this subspace in general do not vanish at site 5. These special properties of the wavefunctions for each band are verified numerically and will be used below to differentiate a topological boundary mode from a non-topological one.

For a 1D periodic system, the Zak phase, as a special form of Berry phase, is the integration of the Berry connection over the Brillouin zone, i.e., $\theta = i \oint dk \langle u(k)|\partial_k u(k)\rangle$, where $|u(k)\rangle$ are the cell periodic eigenstates of the Bloch Hamiltonian.

Since the Bloch wavefunction of the effective SSH model can be expressed as $|\psi_\pm(k)\rangle = \frac{1}{\sqrt{2}}[\pm e^{-i\phi(k)}|1'\rangle + |2'\rangle]$ for the eigen-energies $E = E_\pm(k)$, respectively, we calculate the Zak phase for the $E = E_+(k)$ band as follows:

$$\theta_+ = i\int_{-\pi}^{\pi}\left\langle\psi_+(k)\left|\frac{\partial}{\partial k}\psi_+(k)\right.\right\rangle dk = \frac{i}{2}\int_{-\pi}^{\pi}(e^{i\phi(k)}\langle 1'| + \langle 2'|)\frac{\partial}{\partial k}(e^{-i\phi(k)}|1'\rangle + |2'\rangle)dk =$$

$$\frac{i}{2}\int_{-\pi}^{\pi}e^{i\phi(k)}\langle 1'|\frac{\partial}{\partial k}e^{-i\phi(k)}|1'\rangle dk = \frac{1}{2}\int_{-\pi}^{\pi}\frac{\partial\phi(k)}{\partial k}dk = \frac{\Delta\phi}{2},$$

(10)

where $\Delta\phi$ is the change of $\phi(k)$ when $k$ varies across the Brillouin zone. Similarly, the Zak phase for the $E = E_-(k)$ band is $\theta_- = \theta_+$. Note $\Delta\phi$ in unit of $2\pi$ is the winding number of the Hamiltonian, hence the Zak phase $\theta_\pm$ is $\pi$ times the winding number[36]. If the intra-cell hopping $t$ is larger than the inter-cell hopping $s$, we have $\Delta\phi = 0$, and the winding number is 0 and Zak phases are $\theta_\pm = 0$[24], hence no topological boundary modes are expected. However, if $t < s$, $\Delta\phi = 2\pi$, and the winding number is 1 and Zak phase becomes $\theta_\pm = \pi$, therefore one topological boundary mode is anticipated. If we put two distinct phases $\theta_\pm = 0$ and $\theta_\pm = \pi$ together and form an interface, one topological interface state should be guaranteed. It should be pointed out that in the derivation of Eq. (10) we have used the fact that the coefficients in Eq. (9) are independent of $k$. If not, there will be additional terms in Eq. (10), i.e., $\langle i'|\frac{\partial}{\partial k}|j'\rangle$, $i,j = 1,2$, and the Zak phase may not be quantized. This will be the case when the NNN hopping is introduced into the system and will be discussed in Section IV.B. The quantization of Zak phase in the subspace implies a hidden inversion symmetry in the subsystem. This becomes evident if we consider the subsystem as an effective SSH model with $|1'\rangle$ and $|2'\rangle$ as its two

"atomic" orbitals in a unit cell. Obviously in this subspace, the inversion symmetry is preserved since the Hamiltonian is equivalent to an SSH model.

In the following, we will investigate the topological boundary and interface states and the bulk-boundary correspondence in our system.

## IV. Bulk-boundary correspondence

Since the topological zero-energy boundary and interface states are degenerate with the flat band, in order to investigate such states, we set a non-zero $\epsilon_5$ to move the flat band states away from zero energy. This will not alter the two bands with quantized Zak phases as shown in Fig. 2(d) because the two hybridized "atomic" orbitals of the two bands do not involve the $5^{th}$ site. However, the other three bands will be altered and the flat band now becomes dispersive. For the SSH model, if the system is truncated with a weak bond at the end, i.e., when the intra-cell hopping is smaller than the inter-cell hopping, a boundary mode would occur. The system is characterized by a Zak phase of $\pi$ or winding number of 1, which in turn predicts the existence of a boundary mode. This is a manifestation of the bulk-boundary correspondence. The boundary of the chain can be regarded as an interface between the topologically non-trivial phase and the vacuum, which is topologically trivial.

To study these boundary modes, we truncated our system to a finite number of unit cells with two weak bonds at the ends as shown in Fig. 3(a). In our calculation we choose $t = 0.3$ and $s = 1.7$ and $r_1 = r_2 = 1$ for simplicity. From Eq. (9) we know that the values of inter-chain couplings just determine the relative amplitudes of wavefunctions at the two SSH chains and will not influence the existence of topological boundary modes. We found three zero-energy boundary modes in such system: one at the left boundary and two at the right end. In Fig. 3(b) we plot the amplitudes of a mode decaying from the left boundary in log scale. Since the two zero-energy modes decaying from the right boundary are degenerate, they can be regarded as the linear combinations of one even mode and one odd mode with respect to the wavefunctions on the two identical SSH chains. We show the odd and even modes at the right boundary in Fig. 3(c) and (d), respectively. The decay lengths of the three modes are the same and have the value 0.577 (= $\left|\ln\left(\frac{1.7}{0.3}\right)\right|^{-1}$), which agrees with the complex solutions $k = \pi \pm i \ln(s/t)$ of the SSH model in the $2 \times 2$ block of Eq. (6) at zero energy. The typical wavefunction variations in a unit cell are also plotted as insets in Fig. 3(b)-(d). The wavefunction for the mode at the left (right) boundary is non-vanishing on sites 1 and 3 (2 and 4) in a unit cell. And the left boundary mode is odd with respect to two chains, i.e., the wavefunction has opposite signs on sites 1 and 3, as expected from Eq. (9). Similarly, as shown in the insets of Fig. 3(c) and (d), the odd (even) mode has opposite (same) signs on sites 2 and 4.

Two degenerate zero-energy modes at the right boundary is unexpected since usually the bulk-boundary correspondence predicts one mode at each boundary when the winding number is 1. This indicates that the bulk-boundary correspondence in this semi-topological system is not sufficient to predict all the zero-energy boundary modes.

To understand the existence of the extra zero-energy boundary mode at the right boundary, we recall the previous discussions on the hybridized "atomic" orbitals for the SSH block in Eq. (6). Since these two hybridized "atomic" orbitals are antisymmetric combinations of the "atomic" orbitals of the original two SSH chains, the topological boundary modes predicted by the bulk-boundary correspondence in this space must be odd. Thus, the extra even decaying mode at the right boundary shown in Fig. 3(d) must come from the subspace of the $3 \times 3$ block in which the wavefunctions are of the same sign for the pairs at sites (1,3) and (2,4). Mathematically, the existence of this extra decaying mode is supported by the existence of a complex solution $k = \pi - i\, ln\, s/t$ in the eigen-equation of the $3 \times 3$ block at zero energy. Physically, the existence of three zero-energy boundary modes can be understood by the picture of couplings. Without the coupler chain, each of the two identical SSH chains can support two zero-energy boundary modes when truncated with weak bonds. Thus, there are both even and odd modes at each boundary. With the introduction of the coupler chain as shown in Fig. 3(a), the even mode at the left boundary is shifted to other energy and the odd mode survives as a zero-energy mode by adjusting their relative amplitudes according to Eq. (9). However, the inter-chain couplings which appear at sites 1 and 3 have no effects on the modes at the right boundary as the wavefunctions of these two modes have zero amplitude at sites 1 and 3. Therefore two boundary modes survive at right boundary. Thus, in this semi-topological system, the winding number obtained from a subsystem determine the existence of some but not all zero-energy boundary modes as the rest of the system may support additional boundary modes.

After investigating the boundary modes in the finite structure, we are also interested in studying the interface state formed between topologically trivial and non-trivial phases of our system. According to the bulk structure, there can be four types of interfaces as shown in Fig. 4(a)-(d) depending on how the two bulks are connected. All the four types of structure are terminated on both edges with strong bonds so that no boundary modes would occur at the edges. We call these four types of interface structures as (a) 7-site type, (b) 3-site type, (c) 8-site type, (d) 2-site type, where each number denotes the number of sites involved in each domain wall.

We again call the zero modes with opposite (same) signs in the wavefunction with respect to the two identical SSH chains as odd (even) modes. For each of the four types of interfaces, we find one zero-energy interface state which is odd as expected from the bulk-boundary correspondence. Similar to the case of boundary modes, an additional even zero-energy interface state occurs in the cases of 7-site and 2-site types. Since an interface states can be understood as two decaying modes on each side connected with the correct boundary conditions at the interface, the existence of extra modes for types 7-site and 2-site can be seen from the signs and zeros of the wavefunctions shown in Fig. 4(e) and (f). In each case, there exist two even decaying modes connected at the interface. However, the 3-site and 8-site types do not allow even decaying modes on either side. Again, we have shown that the bulk-boundary correspondence predicts a subset of interface states in such a semi-topological system.

## V. Next-nearest-neighbor hoppings

In the previous sections, we have shown that quantized Zak phases and associated topological boundary and interface states can appear in a system without inversion symmetry due to the presence of a topological subsystem. In addition, we have also shown the existence of non-topological boundary modes coming from the rest of the system. In this section, we will investigate the influence of next-nearest-neighbor hoppings. In Fig. 3(e) we label the next-nearest-neighbor hoppings by $\tau_1, \tau_2, \gamma_1$ and $\gamma_2$. We show below that the system can still be block diagonalized to contain one SSH block, but the Zak phases of the two bands associated with the block are no longer quantized in general except that the four NNN hopping parameters satisfy certain relations. Without loss of generality, we first take $\gamma_1 = \gamma_2 = 0$ for simplicity. Similar to the case with only NN hoppings, the Bloch Hamiltonian $\mathcal{H}_1(k)$ can be written as

$$\mathcal{H}_1(k) = \begin{pmatrix} 0 & t+se^{-ik} & 0 & 0 & r_1 \\ t+se^{ik} & 0 & 0 & 0 & \tau_1 \\ 0 & 0 & 0 & t+se^{-ik} & r_2 \\ 0 & 0 & t+se^{ik} & 0 & \tau_2 \\ r_1 & \tau_1 & r_2 & \tau_2 & \epsilon_5 \end{pmatrix}. \tag{11}$$

We block diagonalize the Hamiltonian into the following form:

$$\mathcal{H}_{1BD}(k) = P^{-1}\mathcal{H}_1(k)P = \begin{pmatrix} 0 & t+se^{-ik} & 0 & 0 & 0 \\ t+se^{ik} & 0 & 0 & 0 & 0 \\ 0 & 0 & 0 & t+se^{-ik} & R_1 \\ 0 & 0 & t+se^{ik} & 0 & R_2^* \\ 0 & 0 & R_1 & R_2 & \epsilon_a \end{pmatrix},$$

(12)

where

$$R_1 = \frac{\lambda+\mu}{\sqrt{2}}, R_2 = \frac{\lambda-\mu}{\sqrt{2}}\rho,$$

$$\lambda = \frac{1}{\sqrt{2}}\Sigma_{i=1,2}\sqrt{(\tau_i + r_i e^{-i\phi(k)})(\tau_i + r_i e^{i\phi(k)})},$$

$$\mu = \frac{1}{\sqrt{2}}\Sigma_{i=1,2}\sqrt{(\tau_i - r_i e^{-i\phi(k)})(\tau_i - r_i e^{i\phi(k)})}. \tag{13}$$

and the unitary matrix $P$ has a complex form and is shown as Eq. (A1) in the appendix A. We note that the existence of a $2 \times 2$ SSH block in Eq. (12) is again a result of rank-nullity theorem. Different from the previous case with only NN hoppings, i.e., Eq. (9), now each of the basis vectors for the $2 \times 2$ SSH block in Eq. (12) is a linear combination of four "atomic" orbitals, i.e., $|1\rangle, |2\rangle, |3\rangle, |4\rangle$, with $k$ dependent coefficients. Because of this complication, the Zak phases of these two bands are not quantized in general. However, if the NNN hopping parameters satisfy the constraint of $\tau_1/r_1 = \tau_2/r_2$, these two bands become topological with quantized Zak phases.

Similar to the calculations in Eq. (10) for the case of NN hoppings, a rigorous proof of the quantization of Zak phases of the two SSH bands $E = E_\pm(k)$ is provided in the Appendix A. It is worth mentioning that when the relation $\tau_1/r_1 = \tau_2/r_2$ does not hold, the two blue bands are still unchanged. However, their Zak phases are not quantized.

The proof shown in the Appendix A can be easily generalized to the case when $\gamma_1 \neq 0$ and $\gamma_2 \neq 0$. In this case the quantization of Zak phase still holds as long as the NNN hopping parameters satisfy the proportional relations

$$\tau_1/\tau_2 = \gamma_1/\gamma_2 = r_1/r_2. \tag{14}$$

In Fig. 3(f) we show the band structure and the associated quantized Zak phases for the parameters $\varepsilon_a = 1, t = 0.7, s = 1.3, r_1 = 0.5, r_2 = 1.2, \tau_1 = 0.05, \tau_2 = 0.12, \gamma_1 = 0.075$ and $\gamma_2 = 0.18$. Similar to the case of NN hoppings, the two blue bands remain intact and have quantized Zak phases. The topological phase transition in the subsystem still occurs at the point $t = s$ where band inversion occurs. The existence of quantized Zak phases in the system is also due to hidden inversion symmetry in the subsystem which is analytically proved in Appendix B.

To discuss the effects of NNN hoppings on the zero-energy boundary and interface states, we first have to point out that such states are not expected when Eq. (14) is not satisfied because there does not exist a subsystem which is topological. This is verified numerically. However, when Eq. (14) is satisfied, we still expect topological boundary modes to occur when the subsystem is in a non-trivial phase. This is also verified numerically. We have studied a system composed of a finite number of unit cells with two boundaries ended with weak bonds, and found two zero-energy odd modes with one at each end, which is predicted by the bulk-boundary correspondence. This is also consistent with the existence of two complex solutions, $k = \pi \pm i \ln\left(\frac{s}{t}\right)$, of the eigen-equation of the $2 \times 2$ block in Eq. (12) at zero energy. Unlike the case of NN hoppings, we do not find an additional zero-energy boundary mode at the right end. This can also be seen from the absence of the complex solution $k = \pi - i \ln\left(\frac{s}{t}\right)$ in the eigen-equation of the $3 \times 3$ block at zero energy. Due to the same reason, the number of interface states is 1 at all the four types of interfaces when NNN hoppings are included. Thus, the bulk-boundary correspondence becomes exact.

## VI. Summary and Conclusions

We want to emphasize that the model we investigated above can be generalized to any number of SSH chains. For example in Fig. 5(a), we show the periodic structure of three identical SSH chains coupled by two coupler chains. There are 8 "atomic" orbitals in a unit cell. Similar to the results found in the model of two coupled SSH chains, the Hamiltonian can be block diagonalized to a subsystem described by a $2 \times 2$ SSH

Hamiltonian. Two bands in this subspace have quantized Zak phases. This topological subsystem is decoupled from the rest of the system which is not topological. A typical band structure of such a system is shown in Fig. 5(b), in which the two bands with quantized Zak phases are plotted in blue and other bands in red. Due to the semi-topological property of the system, zero-energy boundary modes exist at the boundaries of the non-trivial phase and interface states between two topologically distinct phases. For a system with a finite number of unit cells terminated with weak bonds, we find one zero-energy boundary mode at the left boundary and three at the right boundary. Since the system is in a non-trivial phase, the bulk-boundary correspondence predicts one topological boundary mode at each end. This is again due to two additional zero modes originating from non-topological subspace. Similarly for interfaces between two distinct topological phases, there can be one or three zero-energy interface states depending on the detailed interface structures. Same as the case of two coupled SSH chains, the bulk boundary correspondence becomes exact (in the sense that it predicts all the boundary modes) when the next nearest neighbor hoppings are included and the hoppings satisfy the proportional relations.

Now we can generalize the system to $n$ identical SSH chains coupled by $n-1$ coupler chains, which is then a Lieb-like lattice ribbon. In the absence of NNN hoppings, we are guaranteed to have $n-(n-1)=1$ set of SSH bands which are topological. When the system is non-trivial and finite in size, there is one zero-energy boundary mode at the left boundary and $n$ at the right boundary. When the NNN hoppings are considered with hopping parameters between any two adjacent SSH chains satisfying Eq. (14), the topological subspace of the SSH model remains and supports one zero-energy boundary mode at each end when system is non-trivial. There will be no additional zero-energy boundary modes at the right end and bulk-boundary correspondence becomes exact.

In summary, we propose a semi-topological quasi-one-dimensional system consisting of two coupled identical SSH chains. We find that quantized Zak phase can appear for the two SSH bands, although the whole system lacks inversion symmetry. The presence of a topological subsystem is due to a hidden inversion symmetry in a subspace of the whole space. We have also studied the zero-energy boundary and interface states of the system with and without NNN hoppings. In the absence of NNN hoppings, we find an extra non-topological zero-energy boundary mode in addition to two topological boundary modes. Similarly, the number of interface states can be 1 or 2 depending on the detailed domain wall structure, but only one has a topological origin. In the presence of NNN hoppings, the quantization of Zak phases of the two SSH bands still holds as long as NNN hopping parameters satisfy certain proportional relation. Consequently, the topological boundary and interface states remain. However, the additional non-topological zero-energy modes disappear and the bulk-boundary correspondence predicts all the boundary modes. The above results are also true for coupled multiple SSH chains. We believe the interesting division of a system into a topological subsystem and non-topological one is not limited to the model we consider here. Other semi-topological system may also be constructed by using a unit cell containing multiple degrees of freedom with certain symmetries in the Hamiltonian.

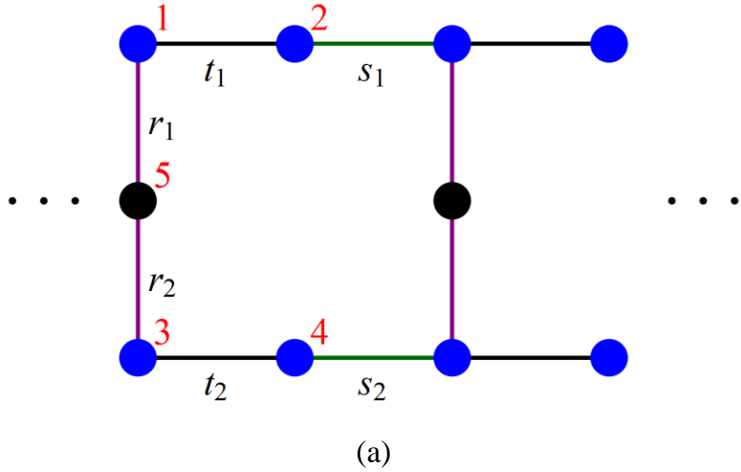

(a)

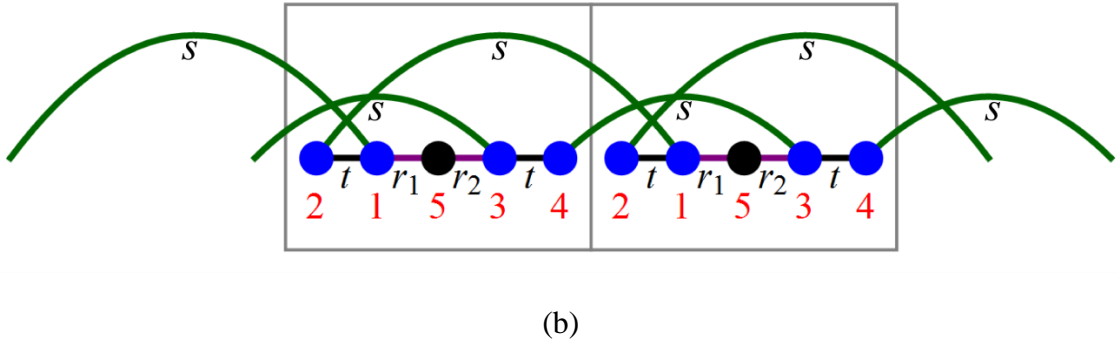

(b)

Fig. 1 (a) Two identical SSH chains coupled by a coupler "atom" per unit cell. The hopping parameters and indices of the "atom" orbitals are labeled. The on-site energies are $\epsilon_1 = \epsilon_2 = \epsilon_3 = \epsilon_4 = 0$ and $\epsilon_5$. (b) The structure in (a) for the case $t_1 = t_2 = t$ and $s_1 = s_2 = s$ is re-drawn as a one-dimensional structure to show the inversion-symmetry-breaking in the system.

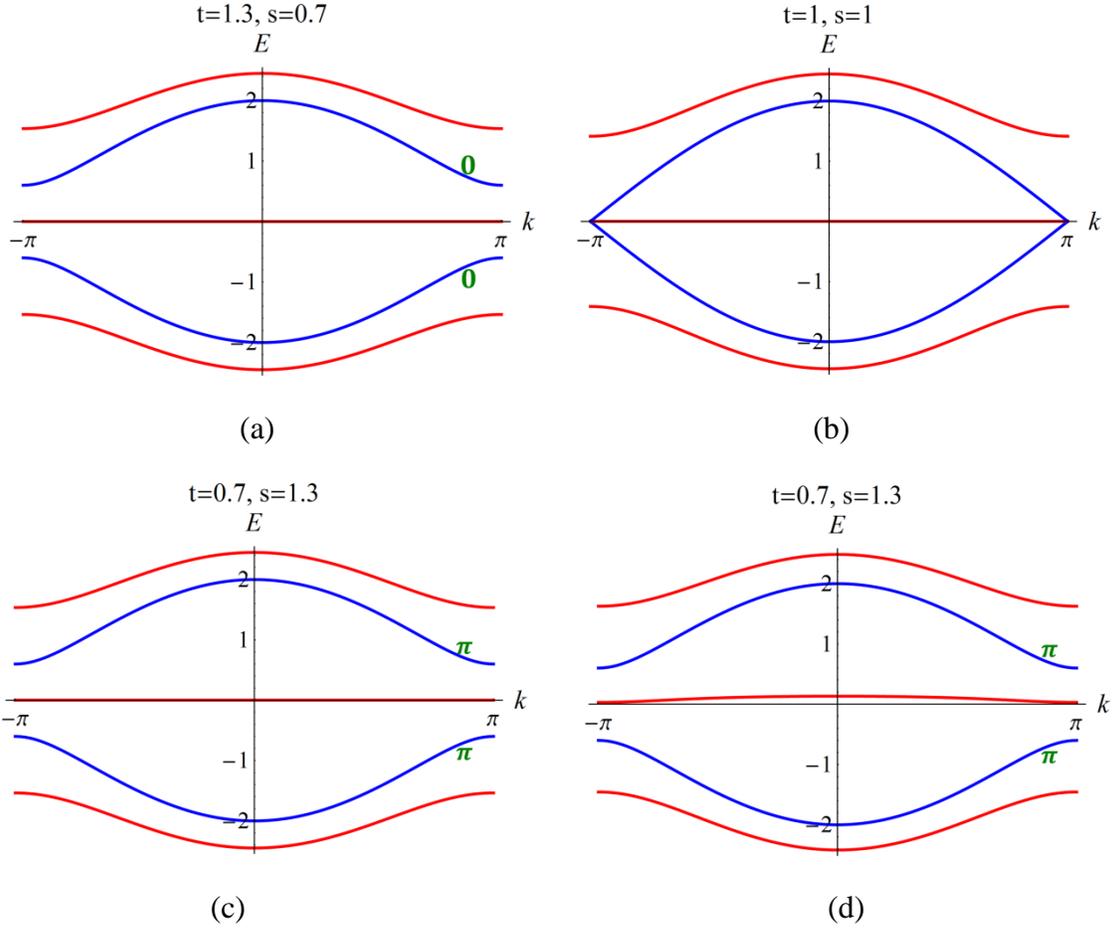

Fig.2 (a), (b) and (c) show the band inversion and quantized Zak phases for the 2$^{nd}$ and 4$^{th}$ bands when hopping parameters are continuously changed from (a) $t > s$ through (b) the transition point $t = s$ to (c) $t < s$. Calculations were done with $r_1 = r_2 = r = 1$ and all the on-site potentials being $0$. Panel (d) shows the band structure with the same parameters as (c) except $\epsilon_5 = 0.2$. The two blue bands remain unchanged with quantized Zak phases. All the red bands are altered and the middle one is no longer flat and moves away from zero energy.

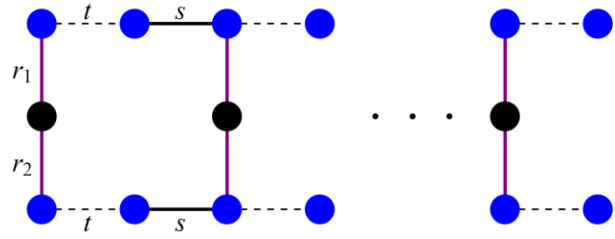

(a)

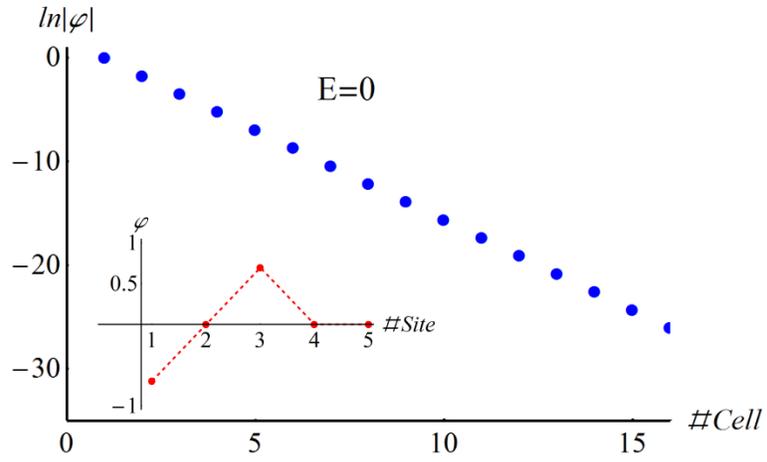

(b)

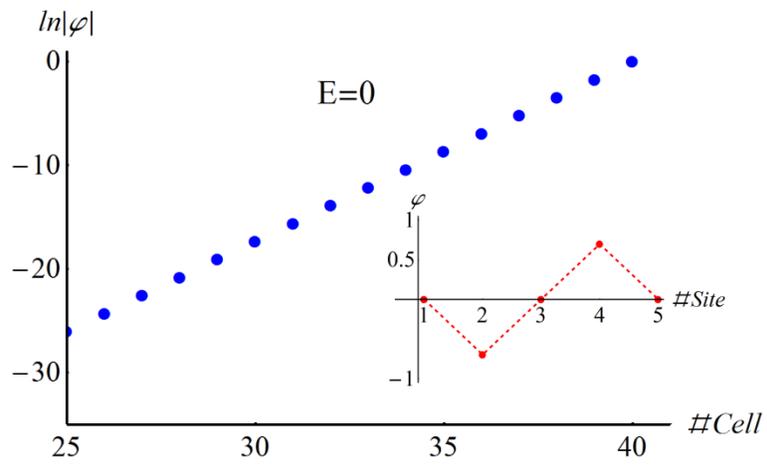

(c)

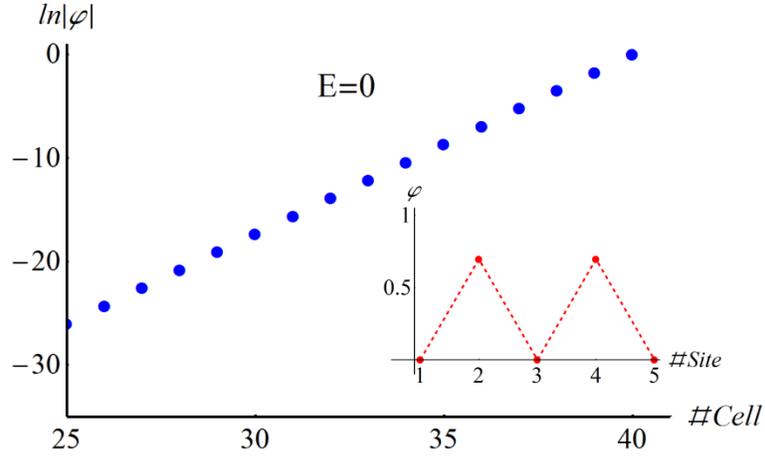

(d)

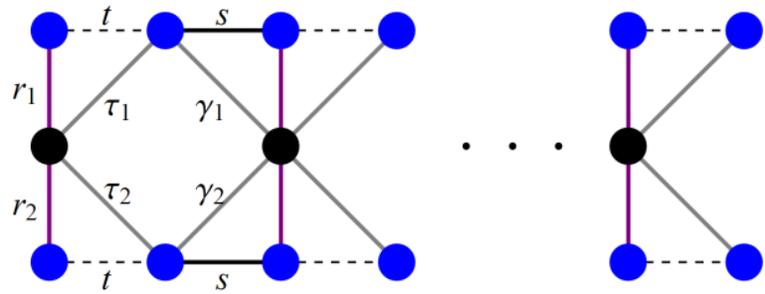

(e)

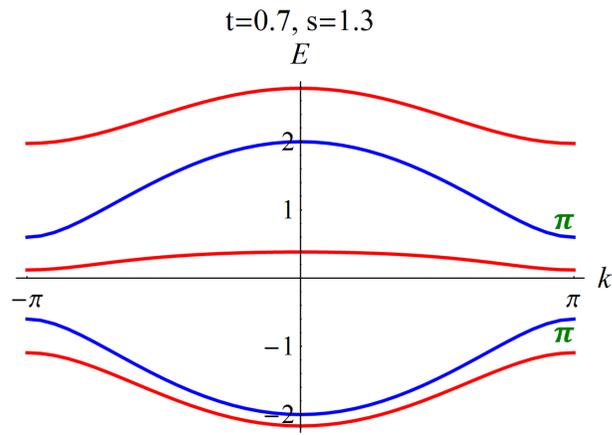

(f)

Fig. 3 (a) shows a finite structure consisting of an integer number of unit cells. The black solid (dashed) lines denote strong (weak) bonds. (b)-(d) show the three topological zero-energy end states for the case $\epsilon_5 = 0.2$. The absolute values of the wavefunctions on the (b) 1$^{st}$ site, (c) 4$^{th}$ site, (d) 4$^{th}$ site in each unit cells are plotted as blue dots against the indices of the unit cells in log scale. Calculations were done with 40 unit cells and the hopping parameters are chosen as $\epsilon_5 = 0.2, t = 0.3, s = 1.7, r = 1$. The insets show the wavefunction variations in the (b) rightmost, (c) rightmost, (d) leftmost unit cell, where 1,2,3,4,5 mean the indices of sites as labeled in Fig. 1(a). (e) shows the introduction of next nearest neighbor hoppings $\tau_1, \tau_2, \gamma_1, \gamma_2$ denoted by grey lines. (f) shows the band structure and the associated quantized Zak phases (labeled in green) in the presence of next nearest neighbor hoppings, when $\frac{r_2}{r_1} = \frac{\tau_2}{\tau_1} = \frac{\gamma_2}{\gamma_1} \equiv \lambda$. The parameters are chosen as $\epsilon_5 = 1, t = 0.7, s = 1.3, \lambda = 2.4, r_1 = 0.5, \tau_1 = 0.05, \gamma_1 = 0.075$.

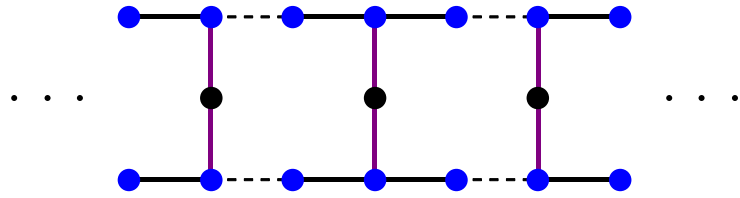

(a)

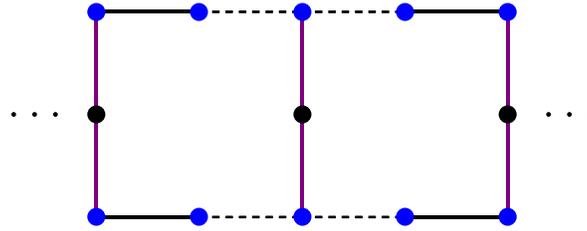

(b)

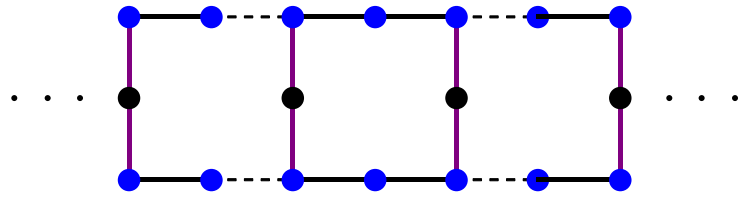

(c)

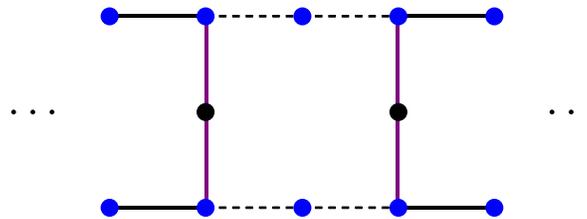

(d)

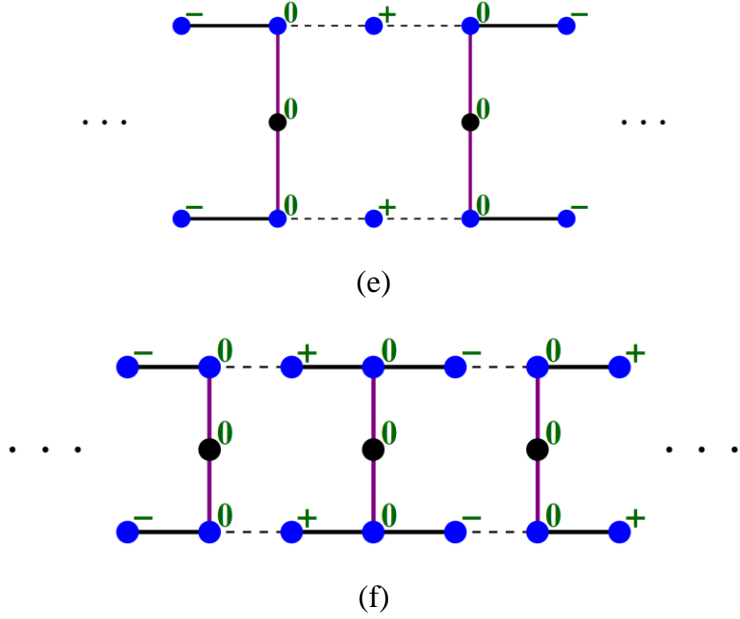

(e)

(f)

Fig. 4 (a)-(d) show four types of domain walls: (a) 7-site type; (b) 3-site type; (c) 8-site type; (d) 2-site type. The black solid (dashed) lines denote strong (weak) bonds in the chain. In the calculations of each structure, the two ends are truncated so that no end states would occur. And the parameters are chosen as $t = 0.2, s = 1.8, r = 0.1$. (e) and (f) depict the two even modes in the cases of 2-site and 7-site types, respectively.

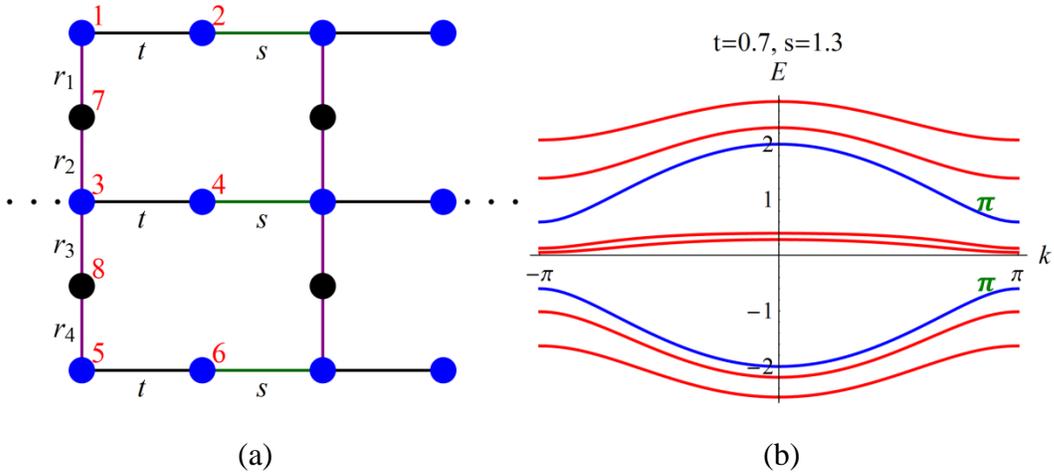

(a)  (b)

Fig. 5 (a) shows the periodic structure of three identical SSH chains coupled by two coupler chains. The hopping parameters and indices of "atomic" orbitals are labeled. (b) shows the band structure of periodic structure in (a). Quantized Zak phases $\theta = \pi$ occur for the two blue bands. The hopping parameters are chosen as $t = 0.7, s = 1.3, r_1 = r_2 = r_3 = r_4 = 1$ and $\epsilon_7 = \epsilon_8 = 0.5$.

# Appendix A

In this appendix, we first show the unitary matrix for the block diagonalization in Eq. (15). Then we prove rigorously the quantization of Zak phases for the two bands $E = E_\pm(k)$ in the subspace of the $2 \times 2$ block in Eq. (14).

The unitary matrix $P$ for the block diagonalization $\mathcal{H}_{1BD}(k) = P^{-1}\mathcal{H}_1(k)P$ of the Hamiltonian $\mathcal{H}_1(k)$ in Eq. (14) for the NNN case is

$$P =$$

$$\frac{1}{2}\begin{pmatrix} (-\frac{a_2+ib_2}{\lambda} - \frac{a_4+ib_4}{\mu}) & (-\frac{a_2+ib_2}{\lambda} + \frac{a_4+ib_4}{\mu})\rho & (\frac{a_1-ib_1}{\lambda} - \frac{a_3-ib_3}{\mu})\rho & (\frac{a_1-ib_1}{\lambda} + \frac{a_3-ib_3}{\mu})\zeta & 0 \\ (-\frac{a_2+ib_2}{\lambda} + \frac{a_4+ib_4}{\mu})\rho^* & (-\frac{a_2+ib_2}{\lambda} - \frac{a_4+ib_4}{\mu}) & (\frac{a_1-ib_1}{\lambda} + \frac{a_3-ib_3}{\mu}) & (\frac{a_1-ib_1}{\lambda} - \frac{a_3-ib_3}{\mu})\rho & 0 \\ (\frac{a_1+ib_1}{\lambda} + \frac{a_3+ib_3}{\mu}) & (\frac{a_1+ib_1}{\lambda} - \frac{a_3+ib_3}{\mu})\rho & (\frac{a_2-ib_2}{\lambda} - \frac{a_4-ib_4}{\mu})\rho & (\frac{a_2-ib_2}{\lambda} + \frac{a_4-ib_4}{\mu})\zeta & 0 \\ (\frac{a_1+ib_1}{\lambda} - \frac{a_3+ib_3}{\mu})\rho^* & (\frac{a_1+ib_1}{\lambda} + \frac{a_3+ib_3}{\mu}) & (\frac{a_2-ib_2}{\lambda} + \frac{a_4-ib_4}{\mu}) & (\frac{a_2-ib_2}{\lambda} - \frac{a_4-ib_4}{\mu})\rho & 0 \\ 0 & 0 & 0 & 0 & 2 \end{pmatrix}.$$

(A1)

where

$$\rho = e^{-i\phi(k)} = \frac{t+se^{-ik}}{E_+}, \quad \zeta = \frac{t+se^{-ik}}{t+se^{ik}} = e^{-2i\phi},$$

$$a_1 = \frac{1}{\sqrt{2}}(\tau_1 + r_1\cos[\phi(k)]), \quad b_1 = -\frac{1}{\sqrt{2}}(r_1\sin[\phi(k)]),$$

$$a_2 = \frac{1}{\sqrt{2}}(\tau_2 + r_2\cos[\phi(k)]), \quad b_2 = -\frac{1}{\sqrt{2}}(r_2\sin[\phi(k)]),$$

$$a_3 = \frac{1}{\sqrt{2}}(\tau_1 - r_1\cos[\phi(k)]), \quad b_3 = \frac{1}{\sqrt{2}}(r_1\sin[\phi(k)]),$$

$$a_4 = \frac{1}{\sqrt{2}}(\tau_2 - r_2\cos[\phi(k)]), \quad b_4 = \frac{1}{\sqrt{2}}(r_2\sin[\phi(k)]),$$

$$\lambda = \frac{1}{\sqrt{2}}\sum_{i=1,2}\sqrt{(\tau_i + r_i e^{-i\phi(k)})(\tau_i + r_i e^{i\phi(k)})},$$

$$\mu = \frac{1}{\sqrt{2}}\sum_{i=1,2}\sqrt{(\tau_i - r_i e^{-i\phi(k)})(\tau_i - r_i e^{i\phi(k)})}.$$

(A2)

We denote the basis vectors for the $2 \times 2$ block in Eq. (15) as $|1'\rangle$ and $|2'\rangle$. Different from the case with only NN hoppings, i.e., Eq. (8), now each of the basis vectors for the $2 \times 2$ SSH block in Eq. (15) is a linear combination of the four "atomic" orbitals $|1\rangle, |2\rangle, |3\rangle, |4\rangle$, with $k$ dependent coefficients

$$|1'\rangle = P_{11}|1\rangle + P_{21}|2\rangle + P_{31}|3\rangle + P_{41}|4\rangle \quad \text{(A3)}$$

$$|2'\rangle = P_{12}|1\rangle + P_{22}|2\rangle + P_{32}|3\rangle + P_{42}|4\rangle \quad \text{(A4)}$$

where the coefficients $P_{ij}$ are elements of the unitary matrix $P$ in Eq. (A1).

In the main text, we stated that the Zak phases for the two bands associated with the $2 \times 2$ block are quantized as long as the hopping parameters satisfy the proportional relation

$$\frac{r_1}{r_2} = \frac{\tau_1}{\tau_2} \equiv \alpha. \quad \text{(A5)}$$

where $\alpha$ is the proportional constant.

Note from the above proportional relation, we can simplify $\lambda$ and $\mu$ in Eq. (A2) to the following form,

$$\lambda = \frac{1}{\sqrt{2}}\sqrt{(\tau_2 + r_2 e^{-i\phi(k)})(\tau_2 + r_2 e^{i\phi(k)})(1 + \alpha^2)},$$

$$\mu = \frac{1}{\sqrt{2}}\sqrt{(\tau_2 - r_2 e^{-i\phi(k)})(\tau_2 - r_2 e^{i\phi(k)})(1 + \alpha^2)}.$$

(A6)

We can also derive the following relations using Eq. (A5)

$$\frac{a_1}{a_2} = \frac{b_1}{b_2} = \frac{a_3}{a_4} = \frac{b_3}{b_4} = \frac{r_1}{r_2} = \alpha. \tag{A7}$$

Immediately we obtain a proportional relation for the coefficients $P_{ij}$ in Eqs. (A3) and (A4),

$$\frac{P_{31}}{P_{11}} = \frac{P_{41}}{P_{21}} = \frac{P_{32}}{P_{12}} = \frac{P_{42}}{P_{22}} = -\frac{r_1}{r_2} = -\alpha. \tag{A8}$$

Then we can simplify the Eqs. (A3) and (A4) to the following form using (A8):

$$|1'\rangle = P_{11}(|1\rangle - \alpha|3\rangle) + P_{21}(|2\rangle - \alpha|4\rangle), \tag{A9}$$

$$|2'\rangle = P_{12}(|1\rangle - \alpha|3\rangle) + P_{22}(|2\rangle - \alpha|4\rangle). \tag{A10}$$

The Zak phase for the positive band $E = E_+(k)$ can be expressed as

$$\theta_+ = i\int_{-\pi}^{\pi}\left\langle\psi_+(k)\left|\frac{\partial}{\partial k}\psi_+(k)\right.\right\rangle dk = \frac{i}{2}\int_{-\pi}^{\pi}(e^{i\phi(k)}\langle 1'| + \langle 2'|)\frac{\partial}{\partial k}(e^{-i\phi(k)}|1'\rangle + |2'\rangle)dk.$$

(A11)

We can write $\theta_+$ as the sum of six parts,

$$\theta_+ = I_1 + I_2 + I_3 + I_4 + I_5 + I_6. \tag{A12}$$

where

$$I_1 = \frac{i}{2}\int_{-\pi}^{\pi}(e^{i\phi(k)}\langle 1'|)|1'\rangle\frac{\partial}{\partial k}(e^{-i\phi(k)})dk,$$

$$I_2 = \frac{i}{2}\int_{-\pi}^{\pi}\langle 1'|\frac{\partial}{\partial k}(|1'\rangle)dk,$$

$$I_3 = \frac{i}{2}\int_{-\pi}^{\pi} e^{i\phi(k)}\langle 1'|\frac{\partial}{\partial k}(|2'\rangle)dk,$$

$$I_4 = \frac{i}{2}\int_{-\pi}^{\pi}\langle 2'||1'\rangle e^{-i\phi(k)}(-i)\frac{\partial\phi(k)}{\partial k}dk,$$

$$I_5 = \frac{i}{2}\int_{-\pi}^{\pi} e^{-i\phi(k)}\langle 2'|\frac{\partial}{\partial k}|1'\rangle dk,$$

$$I_6 = \frac{i}{2}\int_{-\pi}^{\pi}\langle 2'|\frac{\partial}{\partial k}|2'\rangle dk.$$

(A13)

By simple calculations, we obtain

$$I_1 = \frac{i}{2}\int_{-\pi}^{\pi}(-i)\frac{\partial\phi(k)}{\partial k}dk = \frac{1}{2}\int_{-\pi}^{\pi}\frac{\partial\phi(k)}{\partial k}dk = \frac{\Delta\phi}{2}$$

$$I_4 = 0$$

(A14)

Obviously the value of $I_1 = \frac{\Delta\phi}{2}$ is 0 or $\pi$. To calculate the quantity $I_2 + I_3 + I_5 + I_6$ we substitute Eqs. (A9) and (A10) into the expressions of $I_2, I_3, I_5, I_6$ in Eq. (A13) and obtain

$$I_2 = \frac{i}{2}(1+\alpha^2)\int_{-\pi}^{\pi}\left(P_{11}^*\frac{\partial P_{11}}{\partial k} + P_{21}^*\frac{\partial P_{21}}{\partial k}\right)dk,$$

$$I_3 = \frac{i}{2}(1+\alpha^2)\int_{-\pi}^{\pi} e^{i\phi(k)}\left(P_{11}^*\frac{\partial P_{12}}{\partial k} + P_{21}^*\frac{\partial P_{22}}{\partial k}\right)dk,$$

$$I_5 = \frac{i}{2}(1+\alpha^2)\int_{-\pi}^{\pi} e^{-i\phi(k)}\left(P_{12}^*\frac{\partial P_{11}}{\partial k} + P_{22}^*\frac{\partial P_{21}}{\partial k}\right)dk,$$

$$I_6 = \frac{i}{2}(1+\alpha^2)\int_{-\pi}^{\pi}\left(P_{12}^*\frac{\partial P_{12}}{\partial k} + P_{22}^*\frac{\partial P_{22}}{\partial k}\right)dk.$$

(A15)

Thus, we have

$$I_2 + I_3 + I_5 + I_6 = \frac{i}{2}(1+\alpha^2)\int_{-\pi}^{\pi}[J_2(k)+J_3(k)+J_5(k)+J_6(k)]dk \quad (A16)$$

where we denote the four integrands in the four integrals $I_2, I_3, I_5$ and $I_6$ in Eq. (A14) as $J_2(k), J_3(k), J_5(k)$ and $J_6(k)$.

For later convenience, we define

$$A = -\frac{a_2+ib_2}{\lambda}, B = -\frac{a_4+ib_4}{\mu} \quad (A17)$$

Then the four coefficients $P_{11}, P_{12}, P_{21}, P_{22}$ can be written as

$$P_{11} = A + B$$

$$P_{12} = (A-B)\rho$$

$$P_{21} = (A-B)\rho^*$$

$$P_{22} = A + B$$

(A18)

Immediately we can re-write the $J_2(k), J_3(k), J_5(k)$ and $J_6(k)$

$$J_2(k) = P_{11}^*\frac{\partial P_{11}}{\partial k} + P_{21}^*\frac{\partial P_{21}}{\partial k} = (A+B)^*\frac{\partial(A+B)}{\partial k} + (A-B)^*\rho\frac{\partial}{\partial k}[(A-B)\rho^*]$$

$$J_3(k) = P_{11}^*\frac{\partial P_{12}}{\partial k} + P_{21}^*\frac{\partial P_{22}}{\partial k} = (A+B)^*\rho^*\frac{\partial}{\partial k}[(A-B)\rho] + (A-B)^*\frac{\partial}{\partial k}(A+B)$$

$$J_5(k) = P_{12}^*\frac{\partial P_{11}}{\partial k} + P_{22}^*\frac{\partial P_{21}}{\partial k} = (A-B)^*\frac{\partial}{\partial k}(A+B) + (A+B)^*\rho\frac{\partial}{\partial k}[(A-B)\rho^*]$$

$$J_6(k) = P_{12}^*\frac{\partial P_{12}}{\partial k} + P_{22}^*\frac{\partial P_{22}}{\partial k} = (A-B)^*\rho^*\frac{\partial}{\partial k}[(A-B)\rho] + (A+B)^*\frac{\partial(A+B)}{\partial k}$$

We then have

$$J_2(k) + J_3(k) + J_5(k) + J_6(k)$$

$$= 2(A+B)^* \frac{\partial(A+B)}{\partial k} + 2(A-B)^* \frac{\partial}{\partial k}(A+B)$$

$$+ (A-B)^*\rho \frac{\partial}{\partial k}[(A-B)\rho^*] + (A+B)^*\rho^* \frac{\partial}{\partial k}[(A-B)\rho]$$

$$+ (A+B)^*\rho \frac{\partial}{\partial k}[(A-B)\rho^*] + (A-B)^*\rho^* \frac{\partial}{\partial k}[(A-B)\rho]$$

$$= 4A^* \frac{\partial(A+B)}{\partial k} + 2A^*\rho \frac{\partial}{\partial k}[(A-B)\rho^*] + 2A^*\rho^* \frac{\partial}{\partial k}[(A-B)\rho]$$

$$= 4A^* \frac{\partial(A+B)}{\partial k} + 2A^*(A-B)\rho \frac{\partial\rho^*}{\partial k} + 2A^* \frac{\partial}{\partial k}(A-B)$$

$$+ 2A^* \frac{\partial}{\partial k}(A-B) + 2A^*(A-B)\rho^* \frac{\partial\rho}{\partial k}$$

$$= 4A^* \frac{\partial(A+B)}{\partial k} + 4A^* \frac{\partial}{\partial k}(A-B) + 2A^*(A-B)\left[\rho \frac{\partial\rho^*}{\partial k} + \rho^* \frac{\partial\rho}{\partial k}\right]$$

$$= 8A^* \frac{\partial A}{\partial k} + 2A^*(A-B) \frac{\partial}{\partial k}(\rho\rho^*) = 8A^* \frac{\partial A}{\partial k}$$

(A20)

Substituting Eq. (A20) to Eq. (A16) gives

$$I_2 + I_3 + I_5 + I_6 = 4i(1+\alpha^2) \int_{-\pi}^{\pi} A^* \frac{\partial A}{\partial k} dk \tag{A21}$$

We substitute $a_2 + ib_2 = \frac{1}{\sqrt{2}}\left(\tau_2 + r_2 e^{-i\phi(k)}\right)$ and Eq. (6) to Eq. (17) and obtain

$$A = -\frac{a_2 + ib_2}{\lambda} = -\frac{\tau_2 + r_2 e^{-i\phi(k)}}{\sqrt{(\tau_2 + r_2 e^{-i\phi(k)})(\tau_2 + r_2 e^{i\phi(k)})(1+\alpha^2)}}$$

$$= -\frac{1}{\sqrt{(1+\alpha^2)}}\sqrt{\frac{\tau_2 + r_2 e^{-i\phi(k)}}{\tau_2 + r_2 e^{i\phi(k)}}}$$

(A22)

We can then define

$$e^{-i\beta(k)} \equiv \sqrt{\frac{\tau_2 + r_2 e^{-i\phi(k)}}{\tau_2 + r_2 e^{i\phi(k)}}} \equiv Arg[\tau_2 + r_2 e^{-i\phi(k)}] \quad (A23)$$

Finally the Eq. (A21) becomes

$$I_2 + I_3 + I_5 + I_6 = 4i(1+\alpha^2)\int_{-\pi}^{\pi} A^* \frac{\partial A}{\partial k} dk$$

$$= 4i(1+\alpha^2)\frac{1}{1+\alpha^2}\int_{-\pi}^{\pi} e^{i\beta(k)} \frac{\partial}{\partial k} e^{-i\beta(k)} dk$$

$$= 4i\int_{-\pi}^{\pi} e^{i\beta(k)} e^{-i\beta(k)}(-i)\frac{\partial \beta(k)}{\partial k} dk = 4\int_{-\pi}^{\pi} \frac{\partial \beta(k)}{\partial k} dk = 4\Delta\beta$$

(A24)

where $\Delta\beta$ is $2\pi$ when $t < s$ and $\tau_2 < r_2$ and zero otherwise. Substitute Eq. (A14) and (A24) to Eq. (12), we obtain

$$\theta_+ = \frac{\Delta\phi}{2} + 4\Delta\beta \quad (A25)$$

Since Zak phase is a $Z_2$ invariant and can only take values 0 or $\pi$ (modular $2\pi$), the value of $\Delta\beta$ does not affect the Zak phase. Thus, the value of $\theta_+$ is determined solely by the relative magnitudes of the intra and inter cell hoppings in the SSH chains, i.e., $\theta_+ = \frac{\Delta\phi}{2}$, which is 0 when $t > s$ and $\pi$ when $t < s$. The above proof can be easily applied to the Zak phase $\theta_-$ for the negative band $E = E_-(k)$ and generalized to the case when $\gamma_1 \neq 0, \gamma_2 \neq 0$ with the proportional relation $\tau_1/\tau_2 = \gamma_1/\gamma_2 = r_1/r_2$ satisfied.

## Appendix B

We show explicitly below the existence of hidden inversion symmetry in a subspace of the system when NNN hopping parameters satisfy the proportional relation in Eq. (A5).

We noticed that the basis $|1'\rangle$ and $|2'\rangle$ shown in Eqs. (A9) and (A10) in the Appendix A are linear combinations of the four "atomic" orbitals $|1\rangle, |2\rangle, |3\rangle, |4\rangle$, with $k$ dependent coefficients, in which inversion symmetry is not obvious. To show the inversion symmetry hidden in a subspace of our system, we want to find a new basis $|1''\rangle$ and $|2''\rangle$ such that each of them is a simple linear combination of two "atomic" orbitals with constant coefficients, just like the case shown in Eq. (9) for the NN case.

Since the basis vectors $|1'\rangle$ and $|2'\rangle$ in Eqs. (A9) and (A10) are normalized, thus

$$|P_{11}|^2 + |P_{21}|^2 = \frac{1}{1+\alpha^2}. \tag{B1}$$

We have the following relations for the matrix elements of $P$ shown in Eq. (A1),

$$P_{11} = P_{22},$$

$$\frac{P_{21}}{P_{11}} = \frac{P_{41}}{P_{31}} = C(k) \Rightarrow P_{21} = C(k)P_{11},$$

$$\frac{P_{22}}{P_{12}} = \frac{P_{42}}{P_{32}} = -\frac{1}{C^*(k)} \Rightarrow P_{12} = -C^*(k)P_{22},$$

(B2)

where we have defined $C(k) = P_{21}/P_{11}$.

From Eq. (B1) we obtain,

$$|P_{11}|^2(1 + |C(k)|^2) = \frac{1}{1+\alpha^2} \Rightarrow |P_{11}|^2 = \frac{1}{(1+\alpha^2)(1+|C(k)|^2)}.$$

(B3)

By substituting Eq. (B2) into Eqs. (A9) and (A10), we find

$$|1'\rangle = P_{11}[(|1\rangle - \alpha|3\rangle) + C(k)(|2\rangle - \alpha|4\rangle)],$$

$$|2'\rangle = P_{11}[-C^*(k)(|1\rangle - \alpha|3\rangle) + (|2\rangle - \alpha|4\rangle)].$$

(B4)

Observing the form of Eq. (B4), we find that we can linearly combine $|1'\rangle$ and $|2'\rangle$ so that a new basis $|1''\rangle$ and $|2''\rangle$ are formed and each of them consists of only two "atomic" orbitals, namely

$$|1''\rangle = \frac{1}{\sqrt{1+|C(k)|^2}} \frac{|P_{11}|}{P_{11}} \{|1'\rangle - C(k)|2'\rangle\} = \frac{1}{\sqrt{r_1^2+r_2^2}} (r_2|1\rangle - r_1|3\rangle),$$

$$|2''\rangle = \frac{1}{\sqrt{1+|C(k)|^2}} \frac{|P_{11}|}{P_{11}} \{C^*(k)|1'\rangle + |2'\rangle\} = \frac{1}{\sqrt{r_1^2+r_2^2}} (r_2|2\rangle - r_1|4\rangle),$$

(B5)

where we have used the Eqs. (B3) and (B4). Thus, $|1''\rangle$ and $|2''\rangle$ play the role of hybridized "atomic" orbitals, similar to $|1'\rangle$ and $|2'\rangle$ in the case of NN hoppings only.

Since the corresponding unitary transformation $V$ for the change of basis $|i''\rangle = V |i'\rangle, i = 1,2$ is

$$V = \frac{1}{\sqrt{1+|C(k)|^2}} \frac{|P_{11}|}{P_{11}} \begin{pmatrix} 1 & C^*(k) \\ -C(k) & 1 \end{pmatrix}, \tag{B6}$$

we can apply a corresponding unitary transformation to the upper-left 2 x2 block of $\mathcal{H}_{1BD}(k)$ in Eq. (12) which can be denoted by $H_{2\times 2}(k)$. Such a similarity transformation leads to a new block Hamiltonian of the form

$$h_{2\times 2}(k) = V^{-1} H_{2\times 2}(k) V =$$

$$\frac{1}{1+|C|^2} \begin{pmatrix} -C(t+se^{-ik}) - C^*(t+se^{ik}) & (t+se^{-ik}) - C^{*2}(t+se^{ik}) \\ -C^2(t+se^{-ik}) + (t+se^{ik}) & C(t+se^{-ik}) + C^*(t+se^{ik}) \end{pmatrix}.$$

$$\tag{B7}$$

Using the definition of $C(k)$ in Eq. (B2), we find that

$$C(k)(t+se^{-ik}) + C^*(k)(t+se^{ik}) = 0 \tag{B8}$$

and

$$C(-k) = C^*(k). \tag{B9}$$

From Eq. (B8), we immediately have

$$h_{2\times 2}(k) = \frac{1}{1+|C|^2} \begin{pmatrix} 0 & (t+se^{-ik}) - C^{*2}(t+se^{ik}) \\ -C^2(t+se^{-ik}) + (t+se^{ik}) & 0 \end{pmatrix}.$$

We note $h_{2\times2}(k)$ gives the same energy dispersions $E_\pm(k)$ as the SSH Hamiltonian $H_{2\times2}(k)$, since it is related to the SSH Hamiltonian by a similarity transformation. Finally, we show that the new block Hamiltonian $h_{2\times2}(k)$ indeed possesses inversion symmetry, i.e.,

$\sigma_x h_{2\times2}(k) \sigma_x =$

$\frac{1}{1+|C|^2}\begin{pmatrix}0 & 1\\ 1 & 0\end{pmatrix}\begin{pmatrix}0 & (t+se^{-ik})-C^{*2}(t+se^{ik})\\ -C^2(t+se^{-ik})+(t+se^{ik}) & 0\end{pmatrix}\begin{pmatrix}0 & 1\\ 1 & 0\end{pmatrix} =$

$\frac{1}{1+|C|^2}\begin{pmatrix}0 & 1\\ 1 & 0\end{pmatrix}\begin{pmatrix}(t+se^{-ik})-C^{*2}(t+se^{ik}) & 0\\ 0 & -C^2(t+se^{-ik})+(t+se^{ik})\end{pmatrix} =$

$\frac{1}{1+|C|^2}\begin{pmatrix}0 & -C^2(t+se^{-ik})+(t+se^{ik})\\ (t+se^{-ik})-C^{*2}(t+se^{ik}) & 0\end{pmatrix} = h_{2\times2}(-k),$

(B11)

where we used Eq. (B9).

In conclusion, in the new basis $|1''\rangle$ and $|2''\rangle$, the system is equivalent to an inversion symmetric Hamiltonian represented by $h_{2\times2}(k)$ in Eq. (B11), with hybridized "atomic" orbitals shown in Eq. (B5). The ratio between two inter-chain coupling $r_1$ and $r_2$ determines how the "atomic" orbitals are hybridized. Although the whole system with NNN hoppings lacks inversion symmetry, there is one hidden in a subsystem as shown in Eq. (B11).

†Correspondence address: phchan@ust.hk

## Acknowledgments

We thank Professor Kam Tuen Law, Shubo Wang and Meng Xiao for helpful discussions. This work is supported by Research Grants Council, University Grants Committee, Hong Kong (AoE/P-02/12).